\documentclass{elsart}
\usepackage{english}
\usepackage{natbib}
\begin{document}
\runauthor{Stefan Boettcher}
\begin{frontmatter}
\title{Nature's Way of Optimizing}
\author{Stefan Boettcher$^{1,2,3}$ and Allon Percus$^1$}
\address{$^1$Los Alamos National Laboratory, Los Alamos, NM 87545, USA\\
$^2$CTSPS, Clark Atlanta University, Atlanta, GA 30314, USA\\
$^3$Department of Physics, Emory University, Atlanta, GA 30322, USA}
\date{\today}
\begin{abstract}
We propose a general-purpose method for finding high-quality solutions
to hard optimization problems, inspired by self-organizing processes
often found  in nature.  The method, called {\em Extremal
Optimization\/}, successively eliminates extremely undesirable
components of sub-optimal solutions. Drawing upon models used to
simulate far-from-equilibrium dynamics, it complements  approximation
methods inspired by equilibrium statistical physics, such  as
Simulated Annealing. With only one adjustable parameter, its
performance  proves competitive with,  and often superior to, more
elaborate stochastic optimization procedures.   We demonstrate it here
on two classic hard optimization problems: graph  partitioning and the
traveling salesman problem.
\end{abstract}
\end{frontmatter}

In nature, highly specialized, complex structures often emerge when
their  most inefficient variables are selectively driven to
extinction. Evolution, for example, progresses by selecting {\em
against\/} the few most poorly  adapted species, rather than by
expressly breeding those species best adapted to their
environment~\cite{Darwin}. To describe the dynamics  of systems with
emergent complexity, the concept of ``self-organized criticality''
(SOC) has been proposed~\cite{BTW,bakbook}. Models of SOC often rely  on
``extremal'' processes~\cite{PMB}, where the least fit variables are
progressively eliminated. This principle has been applied successfully
in the   Bak-Sneppen model of evolution~\cite{BS,SBFJ}, where a species $i$
is  characterized by a ``fitness'' value $\lambda_i\in[0,1]$, and the
``weakest'' species (smallest $\lambda$) and its closest dependent
species are successively selected for adaptive changes, getting
assigned new (random) fitness values.  Despite its simplicity, the
Bak-Sneppen model reproduces nontrivial features of paleontological
data, including broadly distributed lifetimes of species, large
extinction events and punctuated equilibrium, without the need for
control parameters.  The {\em extremal optimization\/} (EO) method we
propose draws upon the Bak-Sneppen mechanism, yielding a dynamic
optimization procedure free of selection parameters~\cite{GECCO}. Here we
report on the success of this procedure for two generic optimization
problems,  graph partitioning and the traveling salesman problem.

\begin{figure}[t!]\vspace{5.00in}
\includegraphics{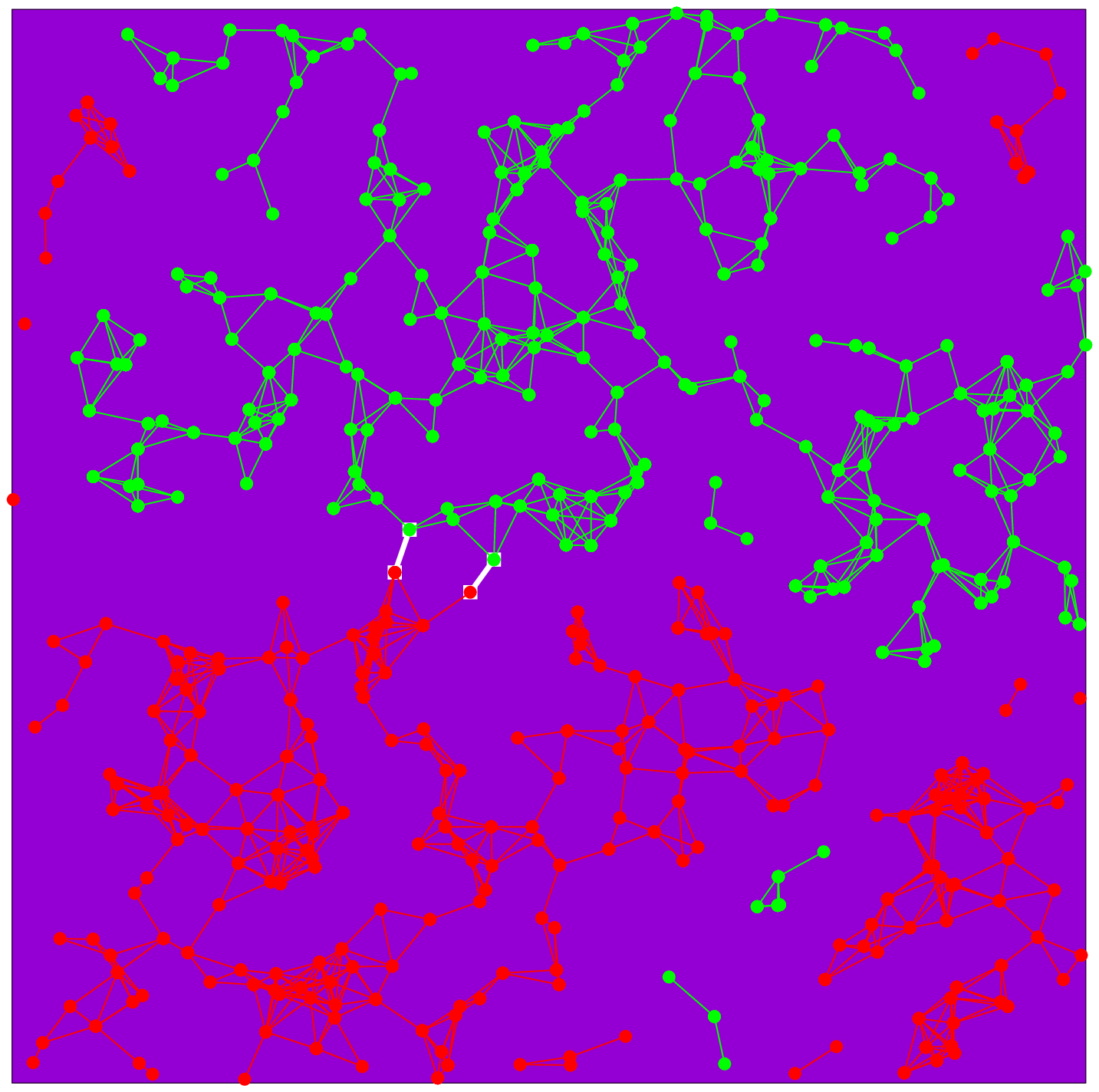}
\caption{Optimal partition of an $N=500$ geometric graph with $C=5$.
Any two points in the unit square are connected
by an edge if their separating distance $d$ satisfies $N\pi d^2<5$.  
The 250 green points make up one subset, and the 250 red points make up the
other.  Over a sample of 30 runs, extremal optimization averaged a cutsize
of 3.7, and eight times found partitions with a cutsize of 2 (shown here in
white).
} 
\label{geograph} 
\end{figure}

In graph (bi-)partitioning, we are given a set of $N$ points, where
$N$ is even, and ``edges'' connecting certain pairs of
points. The problem is to find a way of partitioning the points in two
equal subsets, each of size $N/2$, with a minimal number of edges
cutting across the partition (minimum ``cutsize''). These points, for
instance, could be
positioned randomly in the unit square. A ``geometric'' graph of
average connectivity $C$ would then be formed by connecting any two
points within Euclidean distance $d$, where $N\pi d^2=C$ (see
Fig.~\ref{geograph}).  Constraining the partitioned subsets to be of
fixed (equal) size makes  the solution to this problem particularly
difficult. This geometric problem resembles those found in
VLSI design, concerning the optimal partitioning of gates between 
integrated circuits~\cite{VLSI}. 

Graph partitioning is an {\em NP-hard\/} optimization
problem~\cite{GareyJohnson}: it is believed that for large $N$ the
number of steps necessary for an algorithm to find the {\em exact\/}
optimum must, in general, grow faster than any polynomial in $N$. In
practice, however, the goal is usually to find near-optimal
solutions quickly. Special-purpose heuristics to find approximate
solutions to specific NP-hard problems abound~\cite{AK,JohnsonTSP}.
Alternatively, general-purpose optimization approaches based on
stochastic procedures have been proposed~\cite{Reeves,Osman}.
The most widely applied of these have been physically motivated
methods such as {\em simulated annealing\/}~\cite{SA1,SA2} and
{\em genetic algorithms\/}~\cite{GA,Bounds}.
These procedures, although slower, are applicable to
problems for which no specialized heuristic exists.  EO falls into
the latter category, adaptable to a wide range of combinatorial
optimization problems rather than crafted for a specific application.

Let us illustrate the general form of the EO algorithm by way of the
explicit case of graph bi-partitioning. In close analogy to the
Bak-Sneppen model of SOC~\cite{BS}, the EO algorithm proceeds as follows:
\begin{enumerate}
\item Choose an initial state of the system at will.  In the case of
graph partitioning, this means we choose an initial partition of the
$N$ points into two equal subsets.
\item Rank each variable $i$ of the system according to its fitness value
$\lambda_i$.  For graph partitioning, the variables are the $N$ points,
and we define $\lambda_i$ as follows:
$\lambda_i=g_i/(g_i+b_i)$, where $g_i$ is the number of (good) edges
connecting $i$ to points within the same subset, and $b_i$ is the number 
of (bad) edges connecting $i$ to the other subset. [If point $i$ has no 
connections at all ($g_i=b_i=0$), let $\lambda_i=1$.]  
\item Pick the least fit variable, {\em i.e.\/} the variable with the
smallest $\lambda_i\in [0,1]$, and update it according to some move class.
For graph partitioning, the move class is as follows: the least fit point
(from either subset) is interchanged with a {\em random\/} point from the
other subset, so that each point ends up in the opposite subset from where
it started.
\item Repeat at (2) for a preset number of times. For graph partitioning we 
require $O(N)$ updates.
\end{enumerate}
\noindent The result of an EO run is defined as the best (minimum cutsize)
configuration seen so far.  All that is necessary to keep track of, then,
is the current configuration and the best so far in each run.  

EO, like simulated annealing (SA) and genetic algorithms (GA), is
inspired by  observations of systems in nature.  However, SA emulates
the behavior of frustrated physical systems in thermal equilibrium: if
one couples such a system to a heat bath of adjustable temperature, by
cooling the system slowly one may come close to attaining a state of
minimal energy.  SA accepts or rejects local changes to a
configuration according to the  Metropolis algorithm~\cite{MRRTT} at a
given  temperature, enforcing equilibrium dynamics (``detailed
balance'') and requiring a carefully tuned ``temperature schedule''.
In contrast, EO takes the system far from equilibrium: it applies no
decision criteria, and all new configurations are accepted
indiscriminately.  It may appear that EO's results would resemble an
ineffective random search. But in fact, by persistent selection
against the worst fitnesses, one quickly approaches near-optimal
solutions.  The contrast between EO and genetic algorithms
(GA) is equally pronounced.  GAs keep track of entire ``gene pools''
of states from which to select and ``breed'' an improved generation
of solutions.   EO, on the other hand, operates only with local
updates on a single copy of the system, with improvements achieved
instead by elimination of the bad.

\begin{figure}[t!]\vspace{5.00in}
\includegraphics{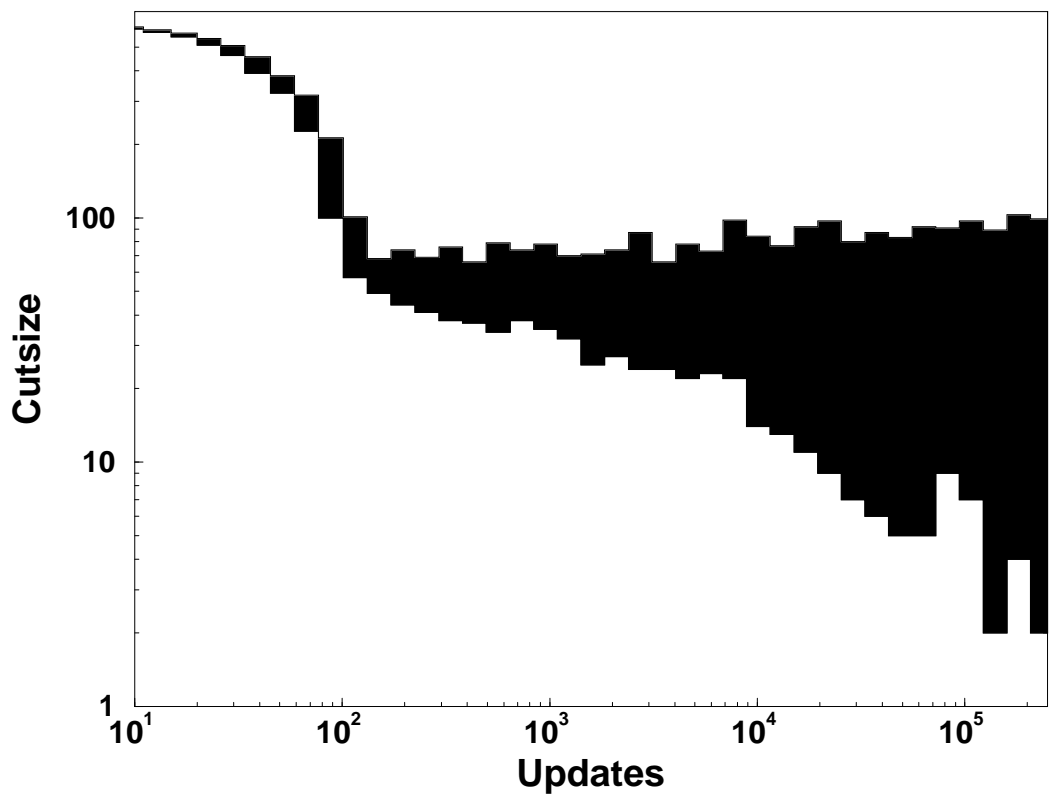}
\caption{Evolution of the cutsize during an extremal optimization run on the
$N=500$ geometric graph with $C=5$ (see Fig.~\protect\ref{geograph}). The
shaded area 
marks the range of cutsizes explored in the respective time bins. The 
best cutsize ever found is 2, which is visited repeatedly in this run.  
In contrast to simulated annealing, which has large fluctuations in 
early stages of the run and then converges much later, extremal optimization 
quickly approaches a stage where broadly distributed fluctuations allow 
it to probe many local optima. In this run, a random initial partition was
used, and the runtime on a 200MHz Pentium was 9sec.
}
\label{runtime} 
\end{figure}

Another important contrast to note is between EO and more
conventional ``greedy'' update strategies.  Methods such as greedy
local search~\cite{Osman} successively update variables so that at
each step, the solution is improved.  This inevitably results in the
system getting stuck in a local optimum, where no further improvements
are possible.  EO, while registering its greatest improvements
towards the beginning of the run, nevertheless exhibits significant
fluctuations throughout, as shown in Fig.~\ref{runtime}.  The result
is that, even at late run-times, EO is able to cross sizable barriers
and access new regions in configuration space.  

There is a closer resemblance between EO and algorithms such as GSAT (for 
satisfiability) that choose, at each update step, the move 
resulting in the best subsequent outcome --- whether or not that outcome 
is an improvement over the current solution~\cite{GSAT}.  Also, versions 
of SA have been proposed~\cite{Greene,Reeves} that enforce equilibrium 
dynamics by ranking local moves according to anticipated outcome, and then
choosing them probabilistically.  Similarly, Tabu Search~\cite{Glover,Reeves}
uses a greedy mechanism based on a ranking of the anticipated
outcome of moves.  But EO, significantly, makes moves using a fitness
that is based not on {\em anticipated\/} outcome but purely on the
{\em current\/} state of each variable.

\begin{figure}[t!]\vspace{5.00in}
\includegraphics{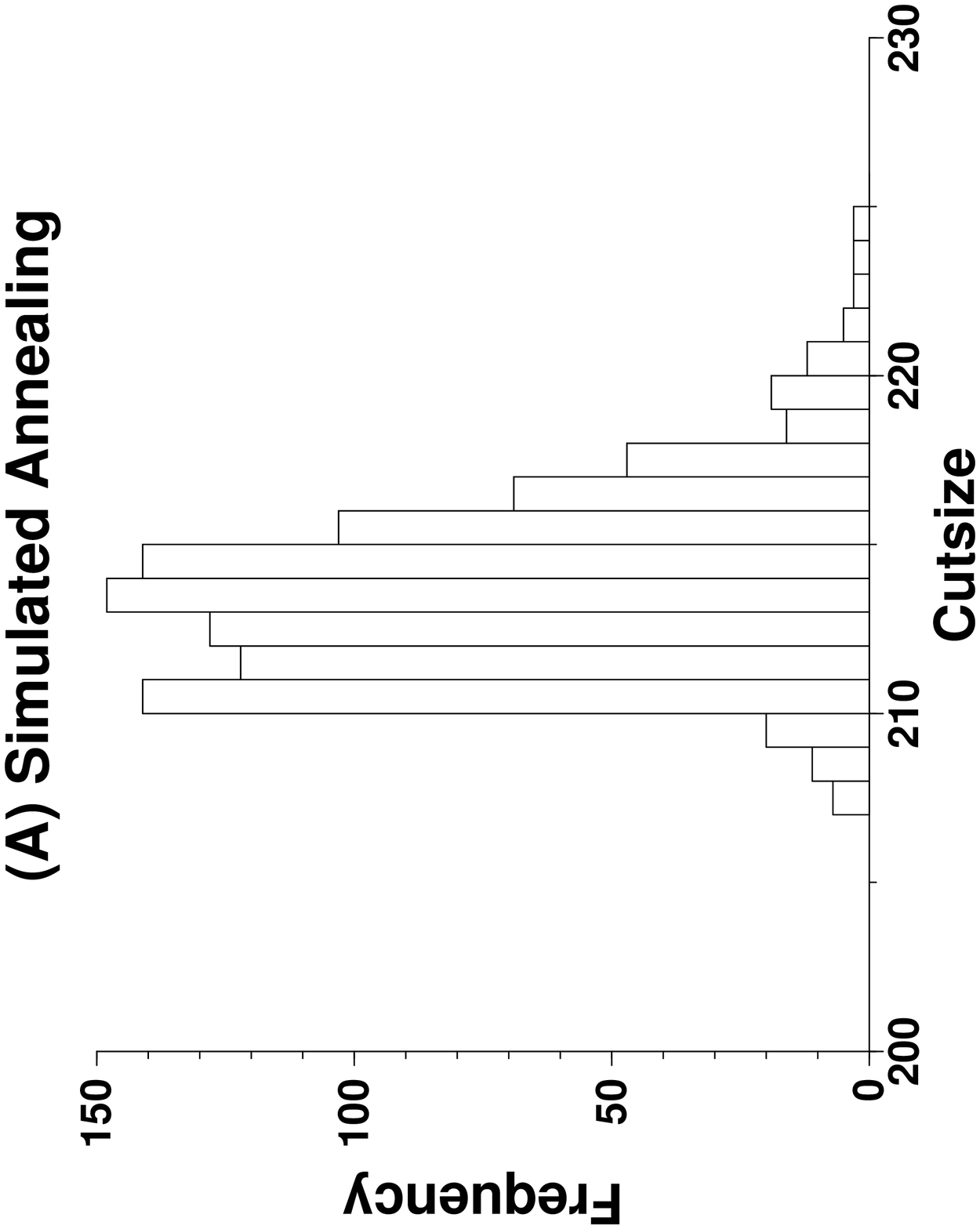}
\includegraphics{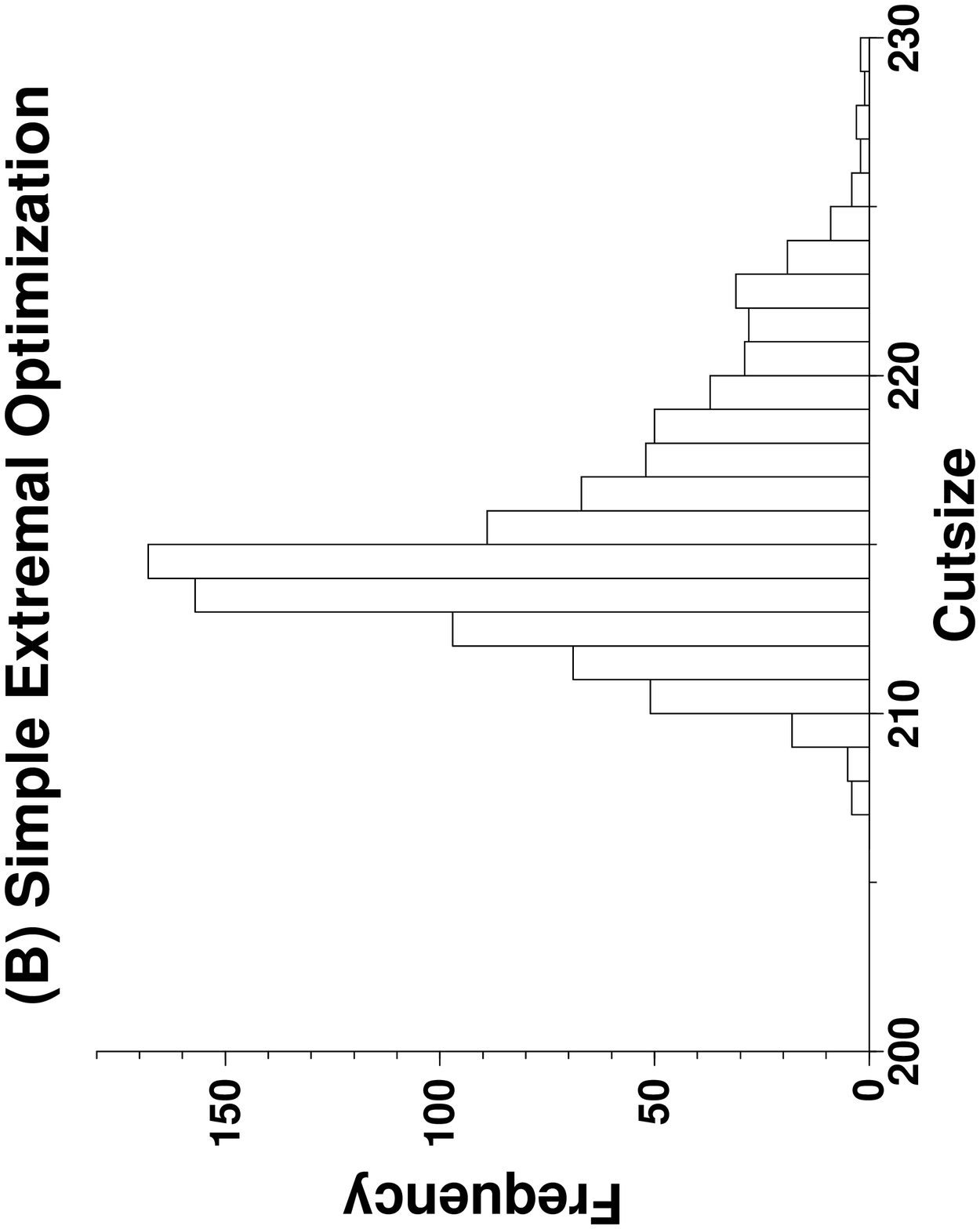}
\includegraphics{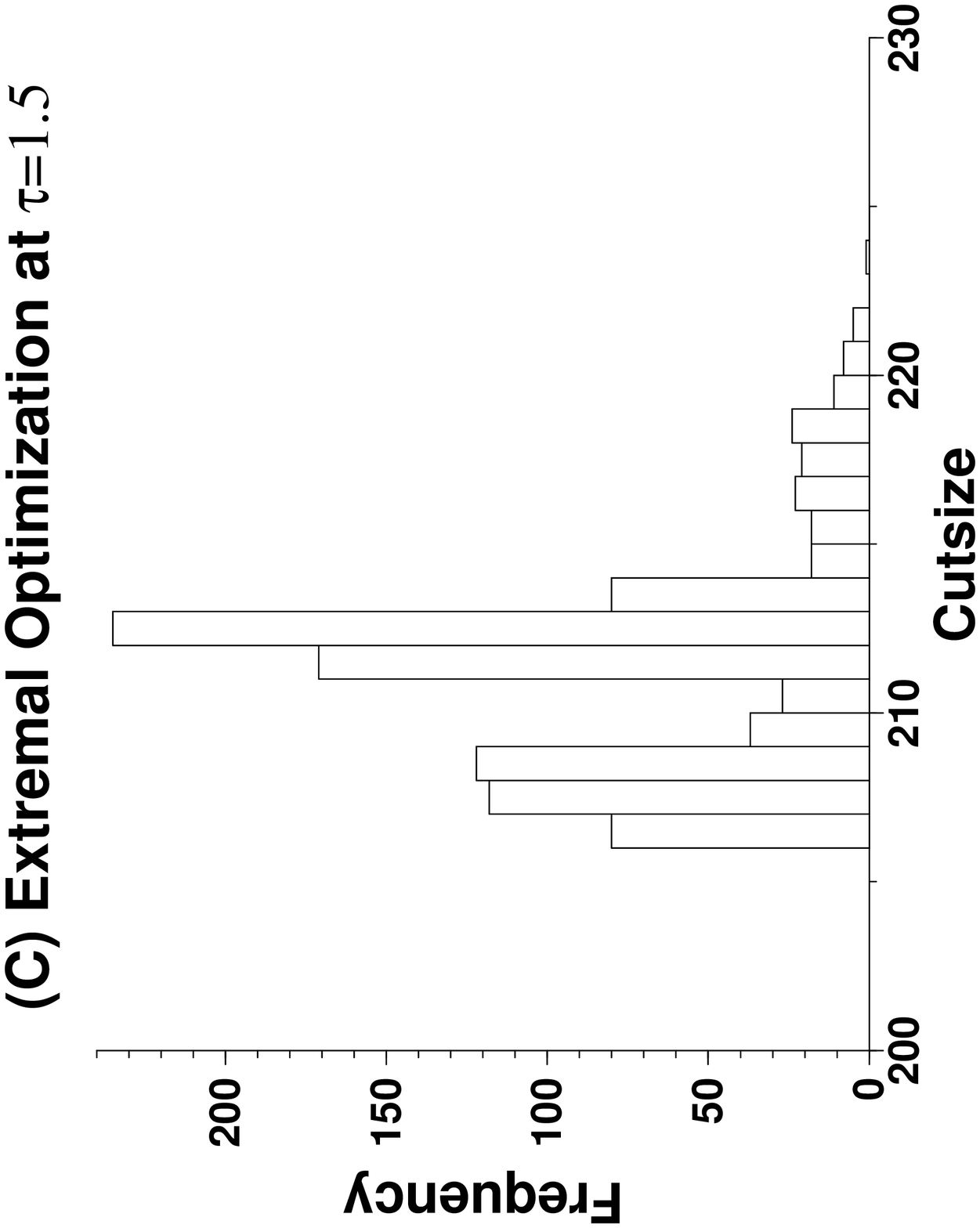}
\caption{Comparison of 1000-run trials using various optimization methods on
$N=500$ random graph with $pN=5$.  
The histograms give, for each method, the frequency
with which a particular cutsize has been obtained during the trial runs.
(A) shows the performance of simulated annealing,
reproducing results given in Ref.~\protect\cite{Johnson}.
(B) shows the results for the basic
implementation of extremal optimization.  (C) shows
the results for extremal optimization using a probability distribution with
$\tau=1.5$.  The best cutsize ever found for this graph is 206.  This result
appeared only once over the 1000 simulated annealing runs, but occurred
80 times over the 1000 extremal optimization runs at $\tau=1.5$.  } 
\label{fig9} 
\end{figure}

Figs.~\ref{fig9}a-b show that the results of EO rival those of a
sophisticated SA algorithm developed for graph
partitioning~\cite{Johnson}.  Further improvements may be obtained
from a slight modification to the EO procedure.  Step (2) of the
algorithm establishes a fitness rank for all points, going from rank
$n=1$ for the worst to rank $n=N$ for the best fitness $\lambda$.
(For points with degenerate values of $\lambda$, the ranks may be
assigned in random order.)  Now relax step (3) so that the points to
be interchanged are {\em both\/} chosen stochastically, from a
probability distribution over the rank order.  This is done in the
following way.  Pick a point having rank $n$ with probability
$P(n)\propto n^{-\tau},~1\leq n\leq N$.  Then pick a second point
using the same process, though restricting ourselves this time to
candidates from the opposite subset.  The choice of a power-law
distribution for $P(n)$ ensures that no regime of fitness gets
excluded from further evolution, since $P(n)$ varies in a gradual,
scale-free manner over rank.  Universally, for a wide range of graphs,
we obtain best results for $\tau\approx 1.2-1.6$.  Fig.~\ref{fig9}c
shows these results for $\tau=1.5$, demonstrating its superior
performance over both SA and the basic EO method.

What is the physical meaning of an optimal value for $\tau$?  If
$\tau$ is too small, we often dislodge already well-adapted points of
high rank: ``good'' results get destroyed too frequently and the
progress of the search becomes undirected.  On the other hand, if
$\tau$ is too large, the process approaches a deterministic local
search (only swapping the lowest-ranked point from each subset) and
gets stuck near a local optimum of poor quality.  At the optimal value
of $\tau$, the more fit variables of the solution are allowed to
survive, without the search being too narrow.  Our numerical studies
have indicated that the best choice for $\tau$ is  closely related to
a transition from ergodic to non-ergodic behavior, with optimal
performance of EO obtained near the edge of ergodicity.  This will be
the subject of future investigation.

To evaluate EO, we applied the algorithm to a testbed of
graphs\footnote{These instances are available via
http://userwww.service.emory.edu/\~{}sboettc/graphs.html} discussed in
Refs.~\cite{Johnson,HL,BM,MF1,MF2}.  The first set of graphs,
originally introduced in Ref.~\cite{Johnson}, consists of eight
geometric and eight ``random'' graphs.  The geometric graphs in the
testbed, labeled ``U$N.C$'', are of sizes $N=500$ and 1000 and
connectivities $C=5$, 10, 20 and 40.  In a random graph, points are
not related by a metric. Instead, any two points are connected with
probability $p$, leading to an average connectivity $C\approx pN$.
The random graphs in the testbed, labeled ``G$Np$'', are of sizes
$N=500$ and 1000 and connectivities $pN=2.5$, 5, 10 and 20.  The best
results reported to date on these graphs have been obtained from
finely-tuned GA implementations~\cite{BM,MF1,MF2}.  EO reproduces most
of these cutsizes, and often at a fraction of the runtime, using
$\tau=1.4$ and 30 runs of $200N$ update steps each. Comparative
results are given in the upper half of Table~\ref{tab1}.

\begin{table}
\caption[t2]{Best cutsizes (and total allowed runtime) for our testbed
of graphs.  Geometric graphs are labeled ``U$N.C$'', and random graphs
are labeled ``G$Np$'' where $C\approx pN$.  GA results are the best
reported in Ref.~\protect\cite{MF1,MF2}, using a 300MHz Pentium.  SA
and EO results are from our runs (SA parameters as determined in
Ref.~\protect\cite{Johnson}), using a 200MHz Pentium.  Comparison data
for three of the large graphs are due to results from heuristics in
Ref.~\cite{HL}, using a 50MHz Sparc20.  }
\bigskip
\begin{tabular}{lr@{}lr@{}lr@{}l|lr@{}lr@{}lr@{}l}
Geom. Graph & \multicolumn{2}{c}{GA} & \multicolumn{2}{c}{SA} &
\multicolumn{2}{c|}{EO} & Rand. Graph & \multicolumn{2}{c}{GA} &
\multicolumn{2}{c}{SA} & \multicolumn{2}{c}{EO} \\ \hline U500.5     &
2 & (13s)  &    4 & (3s)   &    2 & (4s)   & G500.005   &   49 & (60s)
&   51 & (5s)   &   51 & (3s)   \\ U500.10    &   26 & (10s)  &   26 &
(2s)   &   26 & (5s)   & G500.01    &  218 & (60s)  &  219 & (4s)   &
218 & (4s)   \\ U500.20    &  178 & (26s)  &  178 & (1s)   &  178 &
(9s)   & G500.02    &  626 & (60s)  &  628 & (3s)   &  626 & (6s)   \\
U500.40    &  412 & (9s)   &  412 & (.5s)  &  412 & (16s)  & G500.04
& 1744 & (60s)  & 1744 & (3s)   & 1744 & (10s)  \\ U1000.5    &    1 &
(43s)  &    3 & (5s)   &    1 & (8s)   & G1000.0025 &   93 & (120s) &
102 & (9s)   &   95 & (6s)   \\ U1000.10   &   39 & (20s)  &   39 &
(3s)   &   39 & (11s)  & G1000.005  &  445 & (120s) &  451 & (8s)   &
447 & (8s)   \\ U1000.20   &  222 & (37s)  &  222 & (2s)   &  222 &
(18s)  & G1000.01   & 1362 & (120s) & 1366 & (6s)   & 1362 & (12s)  \\
U1000.40   &  737 & (38s)  &  737 & (1s)   &  737 & (33s)  & G1000.02
& 3382 & (120s) & 3386 & (6s)   & 3383 & (20s)  \\ \hline Large Graph
& \multicolumn{2}{c}{GA} & \multicolumn{2}{c}{Ref.~\protect\cite{HL}}
& \multicolumn{2}{c|}{EO} & Large Graph &&& \multicolumn{2}{c}{SA} &
\multicolumn{2}{c}{EO} \\ \hline {\em Hammond\/}    &   90 & (1s)    &
97 & (8s)    &   90 & (42s)   & {\em Nasa1824\/}   &      &         &
739 & (3s)    &  739 & (6s)    \\ \multicolumn{7}{l|}{\quad($N=4720$;
$C=5.8$)}   & \multicolumn{7}{l} {\quad($N=1824$;   $C=20.5$)}  \\
{\em Barth5\/}     &  139 & (44s)   &  146 & (28s)   &  139 & (64s)
& {\em Nasa2146\/}   &      &         &  870 & (2s)    &  870 & (10s)
\\ \multicolumn{7}{l|}{\quad($N=15606$;  $C=5.8$)}  &
\multicolumn{7}{l} {\quad($N=2146$;   $C=32.7$)}  \\ {\em Brack2\/}
&  731 & (255s)  &\multicolumn{2}{c}{---} &  731 & (12s)   & {\em
Nasa4704\/}   &      &         & 1292 & (13s)   & 1292 & (15s)   \\
\multicolumn{7}{l|}{\quad($N=62632$;  $C=11.7$)} & \multicolumn{7}{l}
{\quad($N=4704$;   $C=21.3$)}  \\ {\em Ocean\/}      &  464 & (1200s)
&  499 & (38s)   &  464 & (200s)  & {\em Stufe10\/}    &      &
&  371 & (200s)  &   51 & (180s)  \\
\multicolumn{7}{l|}{\quad($N=143437$; $C=5.7$)} & \multicolumn{7}{l}
{\quad($N=24010$;  $C=3.8$)}  \\
\end{tabular}
\label{tab1}
\end{table}

The next set of graphs in our testbed are of larger size (up to
$N=143$,437).  The lower half of Table~\ref{tab1} summarizes EO's
results on these graphs, again using $\tau=1.4$ and 30 runs.  On each
graph, we used as many update steps as appeared productive for EO to
reliably obtain stable results.  This varied with the particularities
of each graph, from $2N$ to $200N$ (further discussed below), and the
reported runtimes are of course influenced by this.  On the first four
of the large graphs, the best results to date are once again due to
GAs~\cite{MF2}.  EO reproduces all of these cutsizes, displaying an
increasing runtime advantage as $N$ increases.  SA's performance on
the graphs is extremely poor (comparable to its performance on {\em
Stufe10\/}, shown later); we therefore substitute more competitive
results given in Ref.~\cite{HL} using a variety of specialized
heuristics.  EO significantly improves upon these heuristics' results,
though at longer runtimes.  On the final four graphs, for which no GA
results were available, EO matches or dramatically improves upon SA's
cutsizes.  And although the results from the U$N.C$ and G$Np$ graphs
suggest that increasing $C$ slows down EO and speeds up SA, these
results demonstrate that EO's runtime is still nearly competitive with
SA's on the high-connectivity {\em Nasa\/} graphs.

Several factors account for EO's speed.  First of all, we employ a
simple ``greedy'' start to construct the initial partition in step (1),
as follows: pick a point at random, assigning it to one partition, then
take all the points to which it connects, all the points to which
those new points connect, and so on, assigning them all to the same
partition.  When no more connected points are available, construct the
opposite partition by the same means, starting from a new random
(unassigned) point.  Alternate in this way, assigning new points to
one or the other partition, until either one contains $N/2$ points.
This clustering of connected points helps EO converge rapidly, and
instantly eliminates from the running many trivial cases with zero
cutsize.  The procedure is most advantageous for smaller graphs, where
it provides a significant speed-up; that speed-up becomes less
relevant for larger graphs, but can still be productive if the graph
has a distinct non-random structure (this was notably the case for
{\em Brack2\/}).  By contrast, greedy initialization does little to
improve SA: unless the starting temperature is carefully fine-tuned,
any initial advantage is quickly lost in randomization.

Second of all, we use an approximate sorting process in step (2) to
accelerate the algorithm.  At each update step, instead of perfectly
ordering the fitnesses $\lambda_i$ (with runtime factor $CN\log{N}$),
we arrange them on an ordered binary tree called a ``heap''. The
highest level, $l=0$, of this heap is the root of the tree and
consists solely of the poorest fitness.  All other fitnesses are
placed below the root such that a fitness value at the level $l$ is
connected in the tree to a single poorer fitness at level $l-1$, and
to two better fitnesses at level $l+1$.  Due to the binary nature of
the tree, each level has exactly $2^l$ entries, except for the lowest
level $l=[log_2N]$.  We select a level $l$, $0\leq l\leq[log_2N]$,
according to a probability distribution $Q(l)\sim 2^{-(\tau-1)l}$ and
choose one of its $2^l$ entries with equal probability. The rank $n$
distribution of fitnesses thus chosen from the heap roughly
approximates the desired function $P(n)\sim n^{-\tau}$ for a perfectly
ordered list.  The process of resorting the fitnesses in the heap
introduces a runtime factor of only $C\log{N}$ per update step.

A further contributor to EO's speed is the significantly smaller
number of update steps (Fig.~\ref{runtime}) that EO requires compared
to, say, a complete SA temperature schedule.  The quality of our large
$N$ results confirms that $O(N)$ update steps are indeed sufficient
for convergence.  Generally, $200N$ steps were used per run, though in
the case of the {\em Nasa\/} graphs only $30N$ steps were required
for EO to reach its best results, and in the case of the {\em Brack2\/}
graph no more than $2N$ steps were necessary.

In summary, EO appears to be quite successful over a large variety of
graphs.  By comparison, GAs must be finely tuned for each type
of graph in order to be successful, and SA is only useful for
highly-connected graphs; Ref.~\cite{EOperc} demonstrates the dramatic
advantage of EO over SA for sparse graphs.  It is worth noting, though,
that EO's
{\em average\/} performance has been varied.  While on every graph,
the best-found result was obtained at least twice in the 30 runs, the
cutsizes obtained in other runs ranged from a 1\% excess over the best
(on the random graphs) to a 100\% excess or far more (on the others).
For instance, half of the {\em Brack2\/} runs returned cutsizes near
731, but the other half returned cutsizes of above $2000$.  This may
be a product of an unusual structure in this particular graph, as
noted in the discussion above on the initial partition construction.
However, we hope that further insights into EO's performance will be
able to explain these wide fluctuations.

It is also clear that the EO algorithm is applicable to a wide range of
combinatorial optimization problems involving a cost function.  An
example well known to computer scientists is the problem of maximum
satisfiability.  Since one must assign Boolean variables so as to
maximize the number of satisfied clauses, a logical definition of
fitness $\lambda_i$ for a variable $i$ is simply the satisfied
fraction of clauses in which that variable
appears.  Another related problem of great physical interest is the
spin-glass~\cite{MPV}, where spin variables $\sigma_i=\pm1$ on a
lattice are connected via a fixed (``quenched'') network of bonds
$J_{ij}$ randomly assigned values of $+1$ or $-1$ when $i$ and $j$ are
nearest neighbors (and 0 otherwise).  In this system the variables
$\sigma_i$ try to minimize the energy represented by the Hamiltonian
$H=-\sum_{i,j}J_{ij}\sigma_i\sigma_j$.
It is intuitive that the fitness associated with each lattice site
here is the local energy contribution, $\lambda_i={1\over2}
\sigma_i\sum_j J_{ij}\sigma_j$. These applications of EO
have the conceptual advantage that no global constraint needs to be
satisfied, so that on each update a single  variable can be chosen
according to $P(n)\sim n^{-\tau}$; that variable undergoes a
unambiguous flip, affecting the fitnesses of all its neighbors.
We are currently investigating these problems.

In such cases, where the cost can be phrased in terms of a spin
Hamiltonian \cite{MPV}, the implementation of EO is particularly
straightforward.  The concept of fitness, however, is equally
meaningful in any discrete optimization problem whose cost function
can be decomposed into $N$ equivalent degrees of freedom.  Thus, EO
may be applied to many other NP-hard problems, even those where the
choice of quantities for the fitness function, as well as the choice
of elementary move, is less than obvious.  One good example of this
is the traveling salesman problem.  Even so, we find there that EO
presents a challenge to more finely tuned methods.

In the traveling salesman problem (TSP), $N$ points (``cities'') are
given, and every pair of cities $i$ and $j$ is separated by a distance
$d_{ij}$. The problem is to connect the cities using the {\em
shortest\/} closed ``tour'', passing through each city exactly once.
For our purposes, take the $N\times N$ distance matrix $d_{ij}$ to be
symmetric. Its entries could be the Euclidean distances between cities
in a plane --- or alternatively, random numbers drawn from some
distribution, making the problem non-Euclidean.  (The former case
might correspond to a business traveler trying to minimize driving
time; the latter to a traveler trying to minimize expenses on a string
of airline flights, whose prices certainly do not obey triangle
inequalities!)

For the TSP, we implement EO in the following way.  Consider each city
$i$ as a degree of freedom, with a fitness based on the two links
emerging from it.  Ideally, a city would want to be connected to its
first and second nearest neighbor, but is often ``frustrated'' by the
competition of other cities, causing it to be connected instead to
(say) its $\alpha$th and $\beta$th neighbors, $1\leq\alpha,\beta\leq
N-1$.  Let us define the fitness of city $i$ to be
$\lambda_i=3/(\alpha_i+\beta_i)$, so that $\lambda_i=1$ in the ideal
case.

Defining a move class (step (3) in EO's algorithm) is more difficult
for the TSP than for graph partitioning, since the constraint of a
closed tour requires an update procedure that changes several links at
once.  One possibility, used by SA among other local search methods,
is a ``two-change'' rearrangement of a pair of non-adjacent segments
in an existing tour.  There are $O(N^2)$ possible choices for a
two-change.  Most of these, however, lead to even worse results.  For
EO, it would not be sufficient to select two independent cities of
poor fitness from the rank list, as the resulting two-change would
destroy more good links than it creates.  Instead, let us select one
city $i$ according to its fitness rank $n_i$, using the distribution
$P(n)\sim n^{-\tau}$ as before, and eliminate the longer of the two
links emerging from it.  Then, reconnect $i$ to a close neighbor,
using the {\em same\/} distribution function $P(n)$ as for the rank
list of fitnesses, but now applied instead to a rank list of $i$'s
neighbors ($n=1$ for nearest neighbor, $n=2$ for second-nearest
neighbor, and so on).   Finally, to form a valid closed tour, one link
from the new city must be replaced; there is a unique way of doing so.
For the optimal choice of $\tau$, this move class allows us the
opportunity to produce many good neighborhood connections, while
maintaining enough fluctuations to explore the configuration space.

We performed simulations at $N=16$, 32, 64, 128 and 256, in each case
generating ten random instances for both the Euclidean and
non-Euclidean TSP.  The Euclidean case consisted of $N$ points placed
at random in the unit square with periodic boundary conditions; the
non-Euclidean case consisted of a symmetric $N\times N$ distance
matrix with elements drawn randomly from a uniform distribution on the
unit interval.  On each instance we ran both EO and SA from random
initial conditions, selecting for both methods the best of 10 runs.
EO used $\tau=4$ (Eucl.) and $\tau=4.4$ (non-Eucl.), with $16N^2$
update steps\footnote{Given these large values of $\tau$ and
consequently low ranks $n$ chosen, an exact linear sorting of the
fitness list was sufficient, rather than the approximate heap sorting
used for graph partitioning.}.  SA used an annealing schedule with
$\Delta T/T=0.9$ and temperature length $32N^2$.  These parameters
were chosen to give EO and SA virtually equal runtimes.  The results
of the runs are given in Table~\ref{tab2}, along with baseline results
using an exact algorithm~\cite{exact}.

\begin{table}
\caption[t2]{Best tour-lengths found for the Euclidean and the
random-distance TSP.  Results for each value of $N$ are averaged over
10 instances, using on each instance an exact algorithm (except for
$N=256$ Euclidean where none was available), the best-of-ten EO runs
and the best-of-ten SA runs.  Euclidean tour-lengths are rescaled by
$1/\sqrt{N}$.}
\bigskip
\begin{tabular}{rrrr}
$N$ & Exact  & EO$_{10}$  & SA$_{10}$\\  \hline Euclidean:\qquad 16&
0.71453&  0.71453& 0.71453\\ 32& 0.72185&  0.72237& 0.72185\\ 64&
0.72476&  0.72749& 0.72648\\ 128& 0.72024&  0.72792& 0.72395\\ 256&
--- \hspace{.1in} &  0.72707& 0.71854\\ \hline Random Distance:\qquad
16&  1.9368&  1.9368&  1.9368\\ 32&  2.1941&  2.1989&  2.1953\\ 64&
2.0771&  2.0915&  2.1656\\ 128&  2.0097&  2.0728&  2.3451\\ 256&
2.0625&  2.1912&  2.7803
\end{tabular}
\label{tab2}
\end{table}

While the EO results trail those of SA by up to about 1\% in the
Euclidean case, EO significantly outperforms SA for the non-Euclidean
(random distance) TSP.  This may be due to the substantial
configuration space energy barriers exhibited in non-Euclidean
instances; equilibrium methods such as SA get trapped by these
barriers, whereas non-equilibrium methods such as EO do not.
(Interestingly, SA's performance here diminishes rather than improves
when runtimes are increased by using longer temperature schedules!)
For Euclidean instances, the tour lengths found by EO on single runs
were at worst 1\% over the best-of-ten, and the tour lengths found by
SA were at worst 4\% over the best-of-ten; for non-Euclidean
instances, these worst excesses were 5\% (EO) and 10\% (SA).  Finally,
note that one would not expect a general method such as EO to be
competitive here with the more specialized optimization algorithms,
such as Iterated Lin-Kernighan~\cite{CLO,JohnsonILK}, designed
particularly with the TSP in mind.  But remarkably, EO's performance
in both the Euclidean and non-Euclidean cases --- within several
percent of optimality for $N\le 256$ --- places it not far behind the
leading specially-crafted TSP heuristics~\cite{JohnsonTSP}.

Our results therefore indicate that a simple extremal optimization
approach based on self-organizing dynamics can often outperform
state-of-the-art (and far more complicated or finely tuned)
general-purpose algorithms, such as simulated annealing or genetic
algorithms, on hard optimization problems.  Based on its success on
the generic and broadly applicable graph partitioning problem, as well
as on the TSP, we believe the concept will be applicable to numerous
other NP-hard problems.  It is worth stressing that the rank ordering
approach employed by EO is inherently non-equilibrium.  Such an
approach could not, for instance, be used to enhance SA, whose
temperature schedule requires equilibrium conditions.  This rank
ordering serves as a sort of ``memory'', allowing EO to retain
well-adapted pieces of a solution.  In this respect it mirrors one of
the crucial properties noted in the Bak-Sneppen
model~\cite{PRE96,PRL97}.  At the same time, EO maintains enough
flexibility to explore further reaches of the configuration space and
to ``change its mind''.  Its success at this complex task provides
motivation for the use of extremal dynamics to model mechanisms such
as learning, as has been suggested recently to explain the high degree
of adaptation observed in the brain~\cite{bakbrain}.

Thanks to D.~S.~Johnson and O.~Martin for their helpful remarks.

\newpage
\pagestyle{empty}

\newpage
\pagestyle{empty}

\newpage
\pagestyle{empty}

\end{document}